\documentclass[aps,prb,twocolumn,showpacs]{revtex4-1}
\usepackage{graphics}
\usepackage{subfigure}
\usepackage{longtable}
\usepackage{graphicx}
\usepackage{pstricks}
\usepackage{amsmath}
\usepackage{dcolumn}
\usepackage{bm}
\usepackage{pdfpages}

\begin{document}

\title{Origin of the superconducting state in the collapsed tetragonal phase of KFe$_2$As$_2$}

\author{Daniel Guterding}\email{guterding@itp.uni-frankfurt.de}\author{Steffen Backes}\author{Harald O. Jeschke}\author{Roser Valent\'i }
\affiliation{Institut f\"ur Theoretische Physik, Goethe-Universit\"at Frankfurt, 
Max-von-Laue-Stra{\ss}e 1, 60438 Frankfurt am Main, Germany}

\begin{abstract}
  Recently, KFe$_2$As$_2$ was shown to exhibit a structural phase transition 
  from a tetragonal to a collapsed tetragonal phase under applied pressure of 
  about $15~\mathrm{GPa}$. Surprisingly, the collapsed tetragonal phase hosts a 
  superconducting state with $T_c \sim 12~\mathrm{K}$, while the tetragonal phase 
  is a $T_c \leq 3.4~\mathrm{K}$ superconductor. We show that the key difference 
  between the previously known non-superconducting collapsed tetragonal phase in 
  AFe$_2$As$_2$ (A= Ba, Ca, Eu, Sr)  and the superconducting collapsed tetragonal 
  phase in KFe$_2$As$_2$ is the qualitatively distinct electronic structure. While 
  the collapsed phase in the former compounds features only electron pockets at 
  the Brillouin zone boundary and no hole pockets are present in the Brillouin 
  zone center, the collapsed phase in KFe$_2$As$_2$ has almost nested electron 
  and hole pockets. Within a random phase approximation spin fluctuation approach we calculate
  the superconducting order parameter in the collapsed tetragonal phase. 
We propose
  that a Lifshitz transition associated with the structural collapse changes
  the pairing symmetry from $d$-wave (tetragonal) to $s_\pm$ (collapsed tetragonal).
  Our DFT+DMFT calculations show that effects of correlations on 
  the electronic structure of the collapsed tetragonal phase are minimal.
  Finally, we argue that our results are compatible with a change of 
  sign of the Hall coefficient with pressure as observed experimentally.
\end{abstract}


\maketitle

%Introduction section

The family of AFe$_2$As$_2$ (A= Ba, Ca, Eu, K, Sr) superconductors,
also called 122 materials, has been intensively investigated in the
past due to their richness in structural, magnetic and superconducting
phases upon doping or application of pressure~\cite{JohrendtBaK122SC,Kimber2009_NatMat,
Paglione2010_NatPhys,Ca122CoDoping,Ba122CoDoping, Sr122CoDopingRosner}.
   One phase
whose properties have been recently scrutinized at length is the
collapsed tetragonal (CT) phase present in BaFe$_2$As$_2$,
CaFe$_2$As$_2$, EuFe$_2$As$_2$, and SrFe$_2$As$_2$ under pressure and
in CaFe$_2$P$_2$~\cite{Ba122CollapsedTetragonal,
  Ca122CollapsedTetragonalCanfield, Zhang2009_PRB,Dhaka2014_PRB,Sr122CollapsedTetragonalRosner,
  Eu122CollapsedTetragonal, CaFe2P2Coldea, Ca122Profeta}. The structural collapse of
this phase has been shown to be assisted by the formation of As
4$p_z$-As 4$p_z$ bonds between adjacent Fe-As layers giving rise to a
bonding-antibonding splitting of the As $p_z$
bands~\cite{yildirim2009}. It has been argued  that
this phase does not support superconductivity due to the absence of
hole cylinders at the Brillouin zone center and the corresponding
suppression of spin fluctuations~\cite{pratt2009, soh2013, Dhaka2014_PRB}.  However,
recently Ying {\it et al.}~\cite{KFeAsCarrierSwitch} investigated the
hole-doped end member of Ba$_{1-x}$K$_x$Fe$_2$As$_2$, KFe$_2$As$_2$,
under high pressure and observed a boost of the superconducting
critical temperature $T_c$ up to 12 K precisely when the system undergoes
a structural phase transition to a CT phase at a
pressure $P_c \sim$ 15 GPa. These authors attributed this behavior to
possible correlation effects. Moreover, measurements of the Hall
coefficient showed a change from positive to negative sign upon
pressure, indicating that the effective nature of charge carriers
changes from holes to electrons with increasing pressure. Similar experiments
are also reported in Ref. \onlinecite{PaglioneKFe2As2Interband2015}.

\begin{figure}[b]
\includegraphics[width=0.85\linewidth]{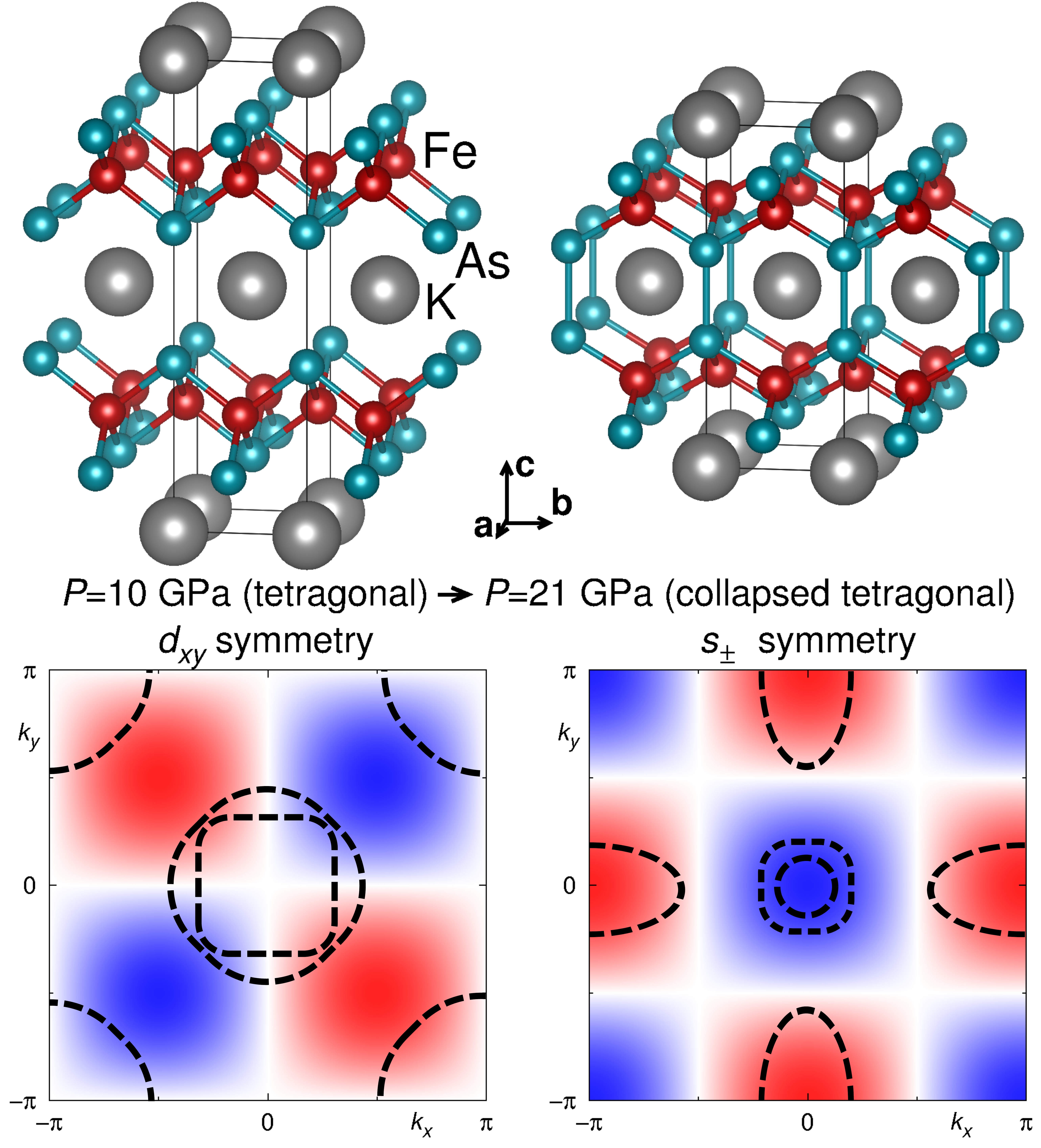}
\caption{(Color online) Crystal structure, schematic Fermi surface (dashed lines) and schematic superconducting gap function (background color)
         of KFe$_2$As$_2$ in the one-Fe Brillouin zone before and after the
         volume collapse. The Lifshitz transition associated with the formation of As 4$p_z$-As 4$p_z$ bonds
         in the CT phase changes the superconducting
 pairing symmetry from $d_{xy}$
         to $s_\pm$.}
\label{fig:pairingtransition}
\end{figure}

KFe$_2$As$_2$ has a few distinct features: at ambient pressure, the
system shows superconductivity at $T_c = 3.4~\mathrm{K}$ and follows a V-shaped
pressure dependence of $T_c$ for moderate pressures with a local
minimum at a pressure of $1.55~\mathrm{GPa}$~\cite{tafti2013}. The origin of
such behavior and the nature of the superconducting pairing symmetry
are still under debate~\cite{okazaki2012, thomale2011, reid2012,
  maiti2011, suzuki2011, tafti2014, tafti2014universal}.  However, it
has been established by a few experimental and theoretical
investigations based on angle-resolved photoemission spectroscopy,
de Haas-van Alphen measurements, and density functional theory combined with dynamical
mean field theory (DFT+DMFT) calculations that
correlation effects crucially influence the behavior of this system at
$P = 0~\mathrm{GPa}$~\cite{sato2009, yoshida2011, yoshida2014,
  terashima2010, terashima2013, haulekotliar2011, KFeAsDMFT,
  kimata2011}. Application of pressure should nevertheless reduce the
relative importance of correlations with respect to the bandwidth
increase. In fact, recent DFT+DMFT studies on CaFe$_2$As$_2$ in the
high-pressure CT phase show that the topology of the
Fermi surface is basically unaffected by
correlations~\cite{CaFeAsCollapsedDiehl, CaFeAsCollapsedHaule}. One
could argue though, that at ambient pressure CaFe$_2$As$_2$ is less
correlated than KFe$_2$As$_2$ and therefore, in KFe$_2$As$_2$
correlation effects may be still significant at finite pressure.

In order to resolve these questions, we performed density functional
theory (DFT) as well as DFT+DMFT
calculations for KFe$_2$As$_2$ in the CT phase. Our
results show that the origin of superconductivity in the collapsed
tetragonal phase in KFe$_2$As$_2$ lies in the qualitative changes in
the electronic structure (Lifshitz
transition) experienced under compression to a collapsed
tetragonal phase and correlations play only a minor role.  Whereas in the
tetragonal phase at $P = 0~\mathrm{GPa}$ KFe$_2$As$_2$ features
predominantly only hole pockets at the Brillouin zone center, at $P
\sim~15~\mathrm{GPa}$ in the CT phase significant
electron pockets emerge at the Brillouin zone boundary, which together
with the hole pockets at the Brillouin zone center favor a superconducting state with
$s_\pm$ symmetry, as we show in our calculations of the
superconducting gap function using the random phase approximation
(RPA) spin fluctuation approach. Moreover, our results
in the tetragonal phase of KFe$_2$As$_2$ at $P = 10~\mathrm{GPa}$
suggest a change of pairing symmetry from $d_{xy}$ (tetragonal) to
$s_\pm$ upon entering the collapsed phase 
(see Fig.~\ref{fig:pairingtransition}).
This scenario is distinct from the
physics of the CT phase in CaFe$_2$As$_2$, where the
hole pockets at the Brillouin zone center are absent.
 For comparison, we will present
the susceptibility of collapsed tetragonal CaFe$_2$As$_2$, which is 
representative for the collapsed phase of AFe$_2$As$_2$ (A= Ba, Ca, Eu, Sr).
 Our findings also suggest an explanation for the change of sign in the
Hall coefficient upon entering the CT phase in
KFe$_2$As$_2$.

%Main section
Density functional theory calculations were carried out using the all-electron 
full-potential local orbital (FPLO)~\cite{FPLOmethod} code. For the 
exchange-correlation functional we use the generalized gradient approximation 
(GGA) by Perdew, Burke, and Ernzerhof~\cite{PerdewBurkeErnzerhof}. All 
calculations were converged on $20 \times 20 \times 20$ k-point grids. 

The structural parameters for the CT phase of KFe$_2$As$_2$ were taken from Ref. 
\onlinecite{KFeAsCarrierSwitch}. We used the data points at $P \approx 
21~\mathrm{GPa}$, deep in the CT phase, where $a = 3.854~\mathrm{\AA}$ and $c = 
9.6~\mathrm{\AA}$. The fractional arsenic $z$-position ($z_\text{As} = 0.36795$) 
was determined {\it ab-initio} via structural relaxation using the FPLO code. We
also performed calculations for the crystal structure of Ref. 
\onlinecite{PaglioneKFe2As2Interband2015}, where a preliminary experimental value
for the arsenic $z$-position was given. The electronic structure is very similar to the 
one reported here.
For the CT phase of CaFe$_2$As$_2$ we used experimental lattice parameters from 
Ref. \onlinecite{Ca122Structure} ($T=40~\mathrm{K}$, $P \approx 
21~\mathrm{GPa}$) and determined the fractional arsenic $z$-position 
($z_\text{As} = 0.37045$) using FPLO. All Fe 3$d$ orbitals are defined in a 
coordinate system rotated by $45^\circ$ around the $z$-axis with respect to the 
conventional $I\,4/mmm$ unit cell.

\begin{figure}[t]
\includegraphics[width=\linewidth]{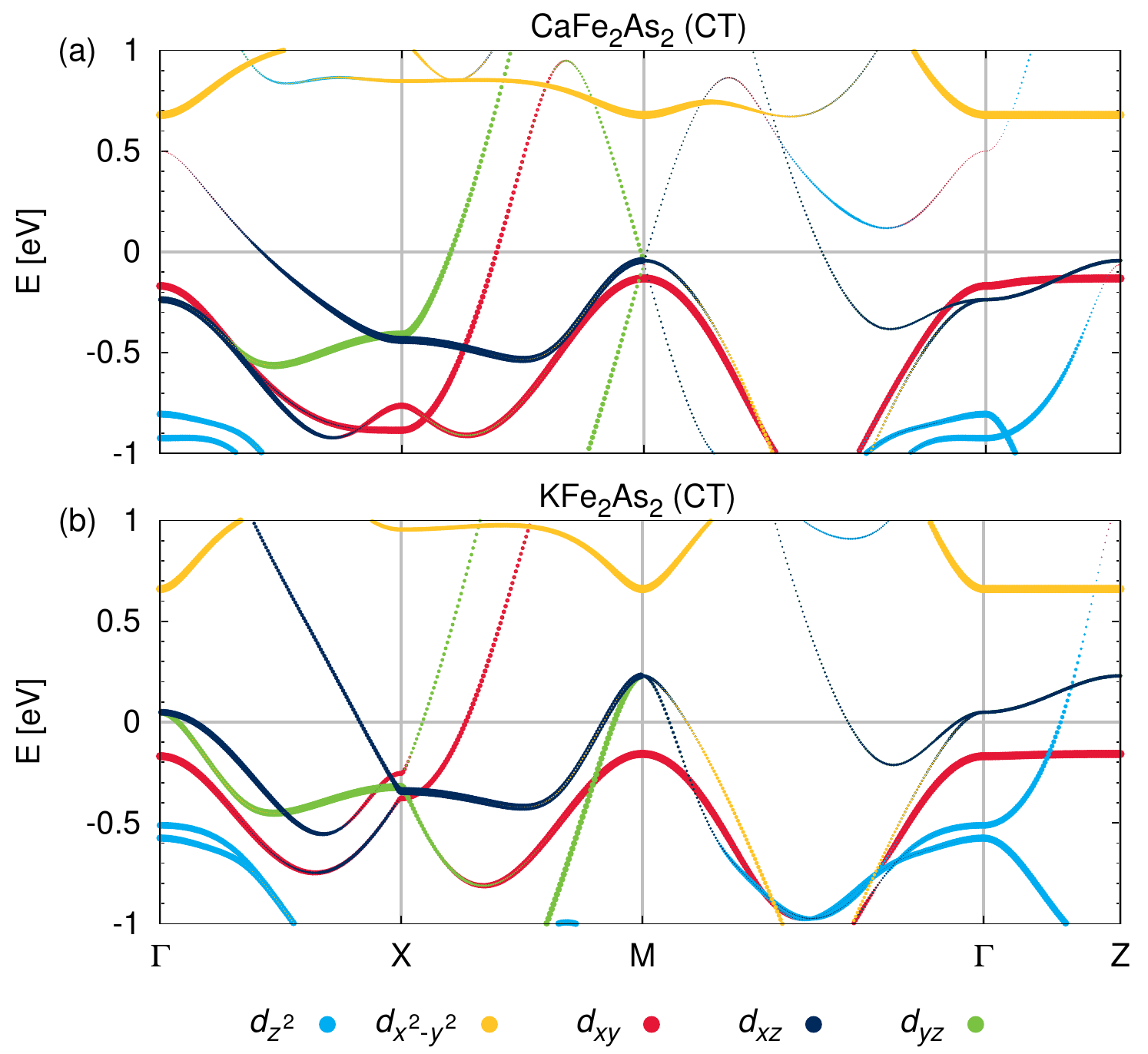}
\caption{(Color online) Electronic bandstructure of the collapsed
  tetragonal phase in (a) CaFe$_2$As$_2$ and (b) KFe$_2$As$_2$. The
  path is chosen in the one-Fe equivalent Brillouin zone. The colors
  indicate the weights of Fe 3$d$ states.}
\label{fig:bandstructure}
\end{figure}

The electronic bandstructure in the collapsed
tetragonal phase of CaFe$_2$As$_2$ and KFe$_2$As$_2$ is shown in
Fig. \ref{fig:bandstructure}. These results already
reveal a striking difference between the CT phases of CaFe$_2$As$_2$
and KFe$_2$As$_2$: while the former does not feature hole bands
crossing the Fermi level at $\Gamma$ and only one band crossing
the Fermi level at M $(\pi, \pi, 0)$, the latter does feature
hole-pockets at both $\Gamma$ and M in the one-Fe equivalent Brillouin
zone. The reason for this difference in electronic structure is that
KFe$_2$As$_2$ is strongly hole-doped compared to CaFe$_2$As$_2$.

\begin{figure}[t]
\includegraphics[width=\linewidth]{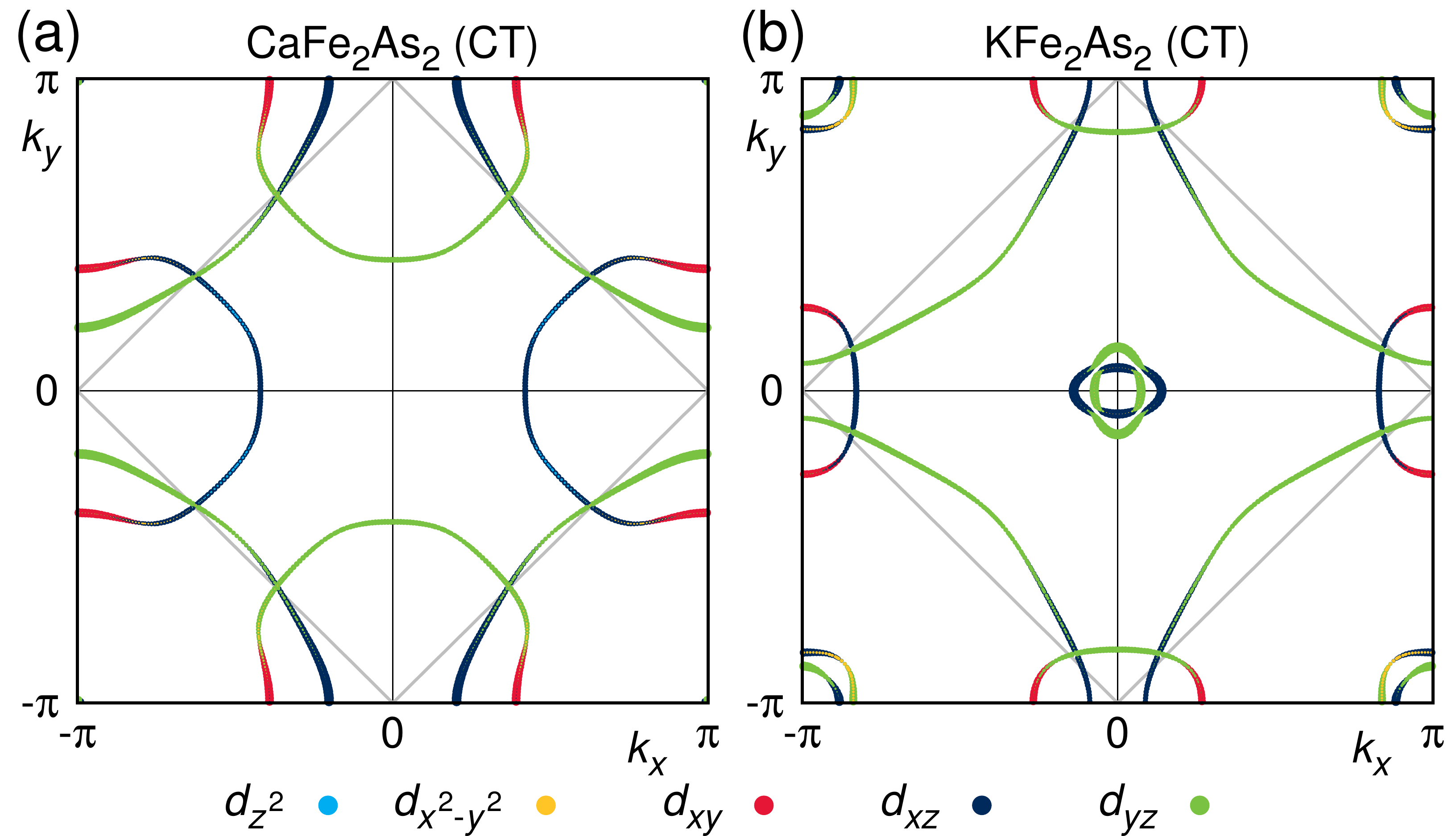}
\caption{(Color online) Fermi surface of the collapsed tetragonal
  phase in (a) CaFe$_2$As$_2$ and (b) KFe$_2$As$_2$ at $k_z = 0$. The
  full plot spans the one-Fe equivalent Brillouin zone, while the area
  enclosed by the grey lines is the two-Fe equivalent Brillouin
  zone. The colors indicate the weights of Fe 3$d$ states.}
\label{fig:fermisurface}
\end{figure}

In Fig. \ref{fig:fermisurface} we show the Fermi surface in the one-Fe equivalent
Brillouin zone at $k_z =0$. In both cases,  the
Fermi surface is  dominated by Fe 3$d_{xz/yz}$
character. The hole cylinders in KFe$_2$As$_2$ span the entire $k_z$ direction of 
the Brillouin zone, while only a small three-dimensional hole-pocket is present in
CaFe$_2$As$_2$ (see Ref. \cite{supplement}).
For KFe$_2$As$_2$, the hole-pockets at M $(\pi, \pi, 0)$ and
the electron pockets at X $(\pi, 0, 0)$ are clearly nested, while no
nesting is observed for CaFe$_2$As$_2$. It is important to note here,
that the folding vector in the 122 family of iron-based
superconductors is $(\pi, \pi, \pi)$, so that the hole-pockets at M
$(\pi, \pi, 0)$ will be located at Z $(0, 0, \pi)$ after unfolding the
bands to the effective one-Fe picture~\cite{TomicUnfolding}.

After qualitatively identifying the difference between the CT phases
of CaFe$_2$As$_2$ and KFe$_2$As$_2$, we calculate the non-interacting
static susceptibility to verify that the better nesting of
KFe$_2$As$_2$ generates stronger spin fluctuations. For that we constructed 16-band tight-binding 
models from the DFT results using projective Wannier functions as implemented in 
FPLO~\cite{FPLOtightbinding}. We keep the Fe 3$d$ and As 4$p$ states, which 
corresponds to an energy window from $-7~\mathrm{eV}$ to $+6~\mathrm{eV}$.
Subsequently, we unfold the 16-band model using our recently developed glide 
reflection unfolding technique~\cite{TomicUnfolding}, which produces an 
effective eight-band model of the three-dimensional one-Fe Brillouin zone. 

We analyse these eight-band models using the 3D version of random
phase approximation (RPA) spin fluctuation
theory~\cite{NJPSpinfluctuationMethod} with a Hamiltonian $H = H_0 +
H_\text{int}$, where $H_0$ is the eight-band tight-binding Hamiltonian
derived from the {\it ab-initio} calculations, while $H_\text{int}$ is
the Hubbard-Hund interaction.
 The arsenic states are
kept in the entire calculation, but interactions are considered only
between Fe 3$d$ states. Further information is given in Ref. \onlinecite{supplement}.

The non-interacting static susceptibility in orbital-space is defined
by Eq.~(\ref{eq:nonintsuscep}), where matrix elements $a^s_{\mu} (\vec
k)$ resulting from the diagonalization of the initial Hamiltonian
$H_0$ connect orbital and band-space denoted by indices $s$ and $\mu$
respectively. The $E_\mu$ are the eigenvalues of $H_0$ and $f(E)$ is
the Fermi function.

\begin{equation}
\begin{array}{rl}
\chi_{st}^{pq} (\vec q) = - \frac{1}{N} \sum \limits_{\vec k, \mu, \nu} & a_\mu^s (\vec k) a_\mu^{p *} (\vec k) a_\nu^q (\vec k + \vec q) a_\nu^{t *} (\vec k + \vec q) \\
& \times \frac{f(E_\nu (\vec k + \vec q)) - f(E_\mu (\vec k))}{E_\nu (\vec k + \vec q) - E_\mu (\vec k)}
\end{array}
\label{eq:nonintsuscep}
\end{equation}

The observable static susceptibility~\cite{supplement} is defined as the sum over all elements 
$\chi_{aa}^{bb}$ of the full tensor
 $\chi (\vec q) = \frac{1}{2} \sum\limits_{a,b} \chi_{aa}^{bb} (\vec q)$.
 
The effective interaction in the singlet
pairing channel is constructed from the static
susceptibility tensor
$\chi_{st}^{pq}$ which measures
strength and wave-vector dependence of spin fluctuations,
 via the multiorbital RPA procedure. Both the original
and effective interaction are discussed, e.g. in
Ref. \onlinecite{HirschfeldReview}. We have
shown previously that our implementation is capable of capturing
effects of fine variations of shape and orbital character of the Fermi
surface~\cite{IntercalateRPA}.

\begin{figure}[t]
\includegraphics[width=\linewidth]{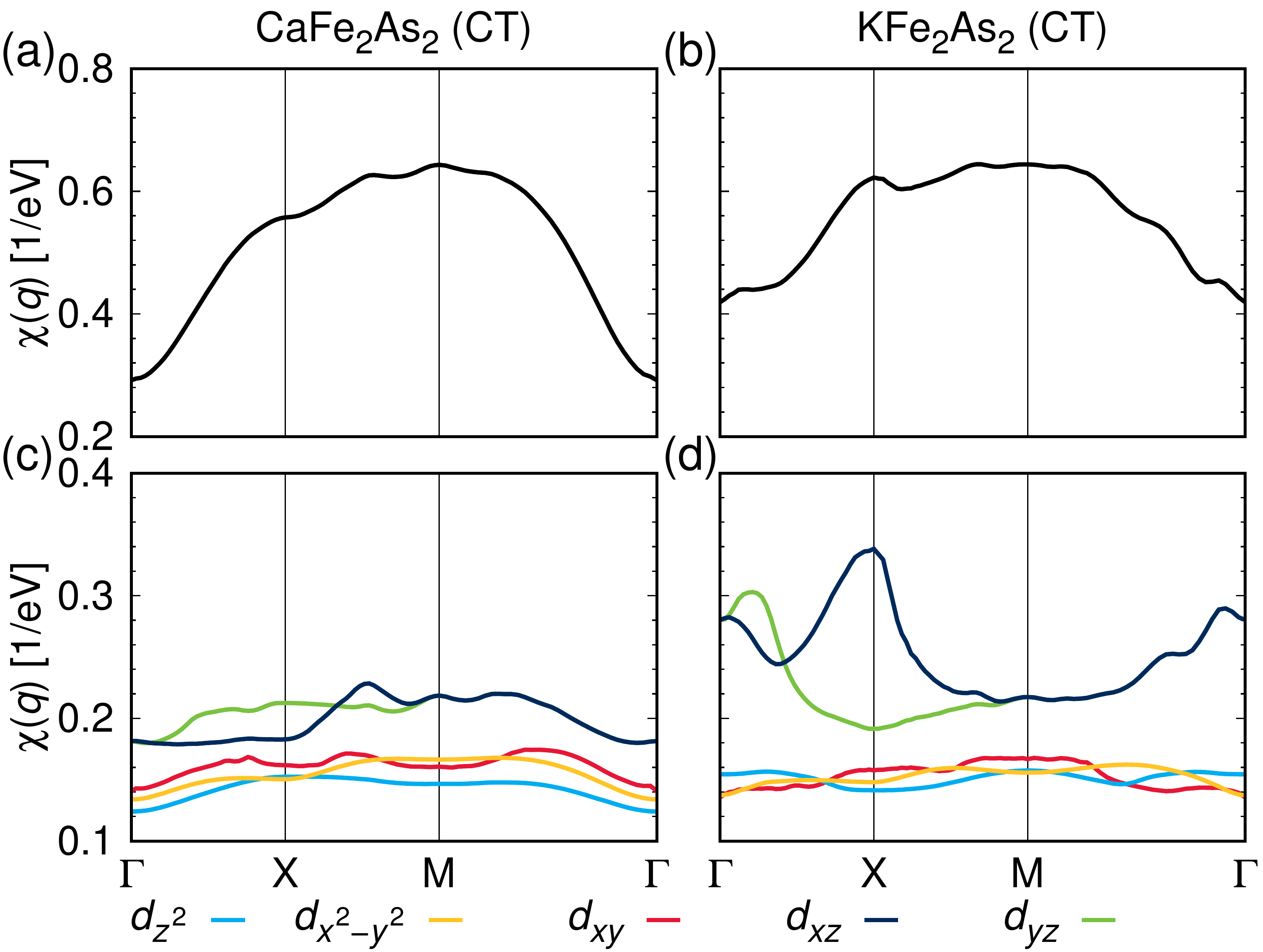}
\caption{(Color online) Summed static susceptibility (top) and its
  diagonal components $\chi_{aa}^{aa}$ (bottom) in the eight-band
  tight-binding model for [(a) and (c)] CaFe$_2$As$_2$ and [(b) and
  (d)] KFe$_2$As$_2$ in the one-Fe Brillouin zone. The colors identify
  the Fe 3$d$ states.}
\label{fig:susceptibility}
\end{figure}

At first glance, the observable static susceptibility displayed in Fig. 
\ref{fig:susceptibility} is comparable for CaFe$_2$As$_2$ and KFe$_2$As$_2$. A 
key difference is however revealed upon investigation of the largest elements, 
i.e. the diagonal entries $\chi_{aa}^{aa}$. These show that in CaFe$_2$As$_2$ 
the susceptibility has broad plateaus, while in KFe$_2$As$_2$ the susceptibility 
has a strong peak at X $(\pi, 0, 0)$ in the one-Fe Brillouin zone, which 
corresponds to the usual $s_\pm$ pairing scenario that relies on electron-hole 
nesting. In CaFe$_2$As$_2$ the pairing interaction is highly frustrated because 
there is no clear peak in favor of one pairing channel.

We have also performed spin-polarized calculations for KFe$_2$As$_2$ at 
$P \approx 21~\mathrm{GPa}$ in order
to confirm the antiferromagnetic instability we find in the linear response
calculations. Out of ferromagnetic, N\'eel and stripe antiferromagnetic order only
the stripe antiferromagnet is stable with small moments of $0.07\mu_B$ on Fe,
in agreement with our calculations for the susceptibility.

\begin{figure}[t]
\includegraphics[width=\linewidth]{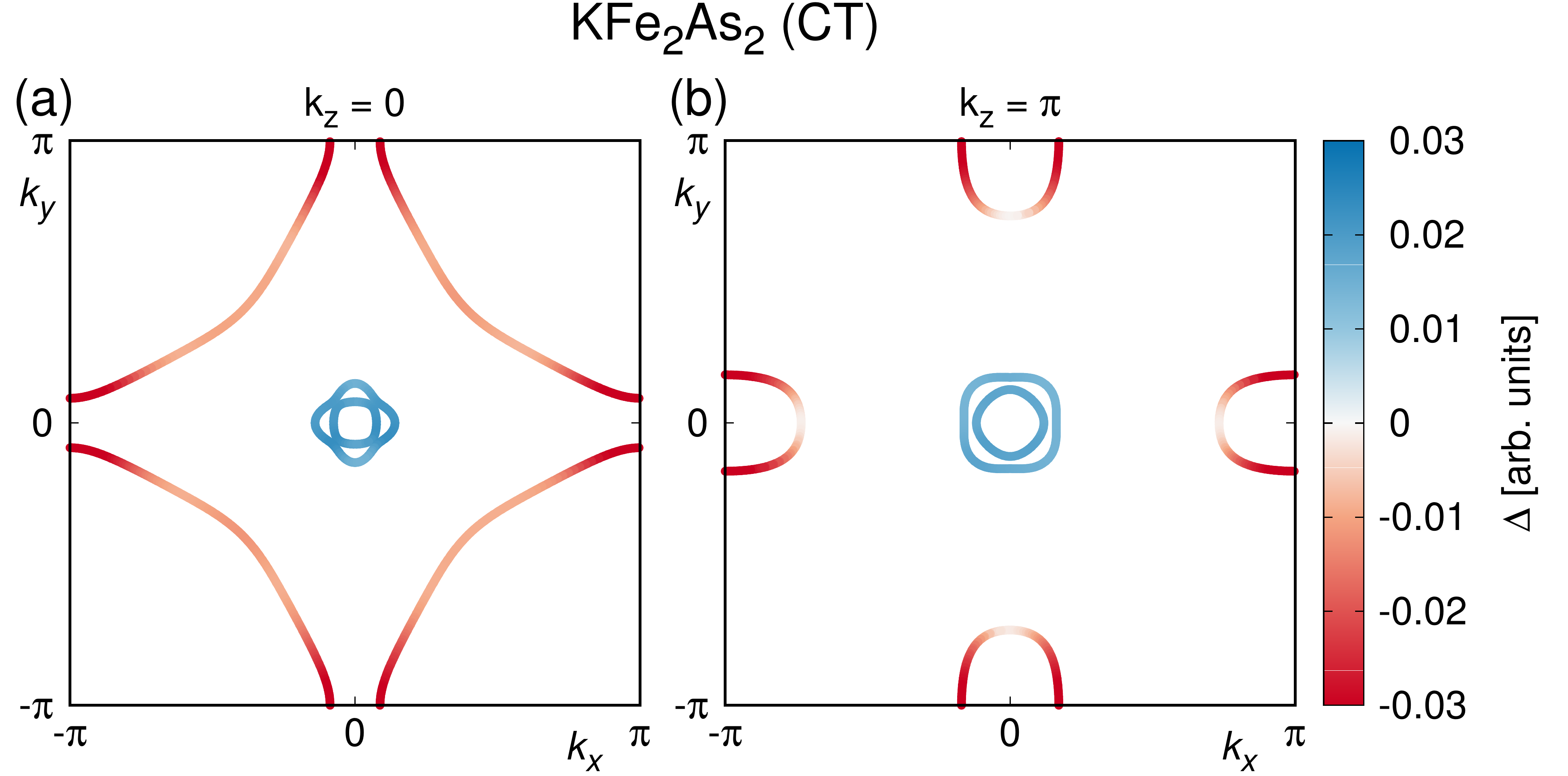}
\caption{(Color online) Leading superconducting gap function ($s_\pm$) of the
  eight-band model in the one-Fe Brillouin zone of KFe$_2$As$_2$ in
  the CT phase at (a) $k_z = 0$ and (b) $k_z = \pi$.}
\label{fig:pairing}
\end{figure}

The leading superconducting gap function of
KFe$_2$As$_2$ in the CT phase is shown in Fig. \ref{fig:pairing}. As expected
from our susceptibility calculations, the pairing symmetry is $s$-wave
with a sign-change between electron and hole-pockets. While the
superconducting gap is nodeless in the $k_z = 0$ plane, the $k_z =
\pi$ plane does show nodes where the orbital character changes from Fe
3$d_{xz/yz}$ to Fe 3$d_{xy}$. Note that this $k_z = \pi$ structure of
the superconducting gap is exactly the same as in the well studied
LaFeAsO compound~\cite{NJPSpinfluctuationMethod}, which shows that the
CT phase of KFe$_2$As$_2$ closely resembles usual iron-based
superconductors although it is much more three-dimensional than, e.g. in LaFeAsO.

We have also calculated the superconducting gap function for KFe$_2$As$_2$ at $P 
= 10$ GPa in the tetragonal phase and find $d_{xy}$ as the leading pairing 
symmetry~\cite{supplement}. The dominant $d_{x^2-y^2}$-solution obtained in 
model calculations based on rigid band shifts~\cite{thomale2011, maiti2011} is 
also present in our calculation, but as a sub-leading solution. Our results 
strongly suggest that the Lifshitz transition, which occurs upon entering the 
collapsed tetragonal phase, changes the symmetry of the superconducting gap 
function from $d$-wave (tetragonal) to $s$-wave (CT) (see Fig. 
\ref{fig:pairingtransition}). The possible simultaneous change of pairing 
symmetry, density of states and $T_c$ potentially opens up different routes to 
understanding their quantitative connection.

In order to estimate the strength of local electronic correlations in
collapsed tetragonal KFe$_2$As$_2$, we performed fully charge
self-consistent DFT+DMFT calculations. We used the same method as
described in Ref.~\onlinecite{KFeAsDMFT}. The DFT calculation was
performed by the \textsc{WIEN2k}~\cite{Blaha01} implementation of the
full-potential linear augmented plane wave (FLAPW) method in the local
density approximation (LDA) with 726 $k$-points in the irreducible Brillouin
zone. We checked that the results of FPLO and \textsc{WIEN2k} agree on the DFT level.
The Bloch wave functions are projected to the localized Fe $3d$
orbitals as described in Refs.~\onlinecite{Aichhorn09,Ferber2014}. The
energy window for projection was chosen from $-7$ to
$+13~\mathrm{eV}$, with the lower boundary lying in a gap in the
density of states. For the solution of the DMFT impurity problem the
continuous-time quantum Monte Carlo method in the hybridization
expansion~\cite{Werner06} as implemented in the
ALPS~\cite{ALPS11,Gull11a} project was employed (see
 Ref. \onlinecite{supplement} for more details).
 The mass-renormalizations are
directly calculated from the analytically continued real part of the
impurity self-energy $\Sigma (\omega)$ via
$\frac{m^*}{m_\text{LDA}} = 1 - \left. \frac{\partial \text{Re}~\Sigma (\omega)}{\partial \omega} \right|_{\omega \to 0}$.

\begin{table}[t]
  \caption{Mass renormalizations $m^*/m_\text{LDA}$ of the Fe 3$d$ orbitals in the collapsed 
    tetragonal phase of KFe$_2$As$_2$ calculated with the LDA+DMFT method.}
\label{tab:massenhancements}
\begin{ruledtabular}
\begin{tabular}{cccc}
$d_{z^2}$ & $d_{x^2-y^2}$ & $d_{xy}$ & $d_{xz/yz}$ \\
1.318 & 1.309 & 1.319 & 1.445 \\
\end{tabular}
\end{ruledtabular}
\end{table}

Table \ref{tab:massenhancements} displays the orbital-resolved
mass-renormalizations $m^*/m_\text{LDA}$ for KFe$_2$As$_2$ in the collapsed
tetragonal phase. The obtained values show
that local electronic correlations in the CT phases
of KFe$_2$As$_2$ and CaFe$_2$As$_2$~\cite{CaFeAsCollapsedDiehl,
  CaFeAsCollapsedHaule} are comparable. As in CaFe$_2$As$_2$, the effects of local
electronic correlations on the Fermi surface are negligible (see Ref. \onlinecite{supplement}). 
The higher $T_c$ of the collapsed phase in absence of strong correlations
raises the question how important strong correlations are in general
for iron-based superconductivity. This issue demands further
investigation.

Finally, the change of dominant charge carriers from hole to
electron-like states measured in the Hall-coefficient under
pressure~\cite{KFeAsCarrierSwitch} is naturally explained from our
calculated Fermi surfaces. While KFe$_2$As$_2$ is known to show only
hole-pockets at zero pressure, the CT phase features
also large electron pockets. On a small fraction of these electron
pockets, the dominating orbital character is Fe 3$d_{xy}$ (Fig.
\ref{fig:fermisurface}). It was shown in
Ref. \onlinecite{HirschfeldHallEffect} that quasiparticle lifetimes on
the Fermi surface can be very anisotropic and long-lived states are
favored where marginal orbital characters appear. As Fe 3$d_{xy}$
character is only present on the electron pockets, these states
contribute significantly to transport and are responsible for the 
negative sign of the Hall coefficient.

%Summary section
In summary, we have shown that the electronic structure of the collapsed tetragonal phase of 
KFe$_2$As$_2$ qualitatively differs from that of other known collapsed 
materials. Upon entering the CT phase, the Fermi surface of 
KFe$_2$As$_2$ undergoes a Lifshitz transition with electron pockets appearing at 
the Brillouin zone boundary, which are nested with the hole pockets at the 
Brillouin zone center. Thus, the spin fluctuations in collapsed tetragonal 
KFe$_2$As$_2$ resemble those of other iron-based superconductors in 
non-collapsed phases and   the superconducting gap function assumes the 
well-known $s_\pm$ symmetry. This is in contrast to other known materials in the 
CT phase, like CaFe$_2$As$_2$, where hole pockets at the 
Brillouin zone center are absent and no superconductivity is favored. Based on 
our LDA+DMFT calculations, the CT phase of KFe$_2$As$_2$ is 
significantly less correlated than the tetragonal phase, and mass enhancements 
are comparable to the CT phase of CaFe$_2$As$_2$.  Finally, we suggest that the 
change of dominant charge carriers from hole to electron-like can be explained 
from anisotropic quasiparticle lifetimes.

%Acknowledgements
\begin{acknowledgments}
  We thank the Deutsche Forschungsgemeinschaft for financial support
  through Grant No. SPP 1458.
\end{acknowledgments}

\bibliographystyle{apsrev4-1}

\clearpage


\section*{Supplementary material}

# Origin of the superconducting state in the collapsed tetragonal phase of KFe$_2$As$_2$: Supplemental Information

Daniel Guterding,* Steffen Backes, Harald O. Jeschke, and Roser Valentí

*Institut für Theoretische Physik, Goethe-Universität Frankfurt,*
*Max-von-Laue-Straße 1, 60438 Frankfurt am Main, Germany*

## I. DETAILS OF THE SUSCEPTIBILITY AND PAIRING CALCULATION

For calculating the susceptibility we used $30 \times 30 \times 10$ k-point grids and an inverse temperature of $\beta = 40\,\mathrm{eV}^{-1}$. To calculate the pairing interaction, the susceptibility is needed at k-vectors that do in general not lie on a grid. Those susceptibility values are obtained using trilinear interpolation of the gridded data.

The pairing interaction is constructed using $\sim 800$ points on the three-dimensional Fermi surface. Consequently, the solution of the gap equation is available only on those points scattered in three-dimensional space. In order to obtain a graphical representation of the gap function on two-dimensional cuts through the Brillouin zone, we used multiscale radial basis function interpolation as implemented in ALGLIB (http://www.alglib.net).

The Hubbard-Hund interaction $H_{\mathrm{int}}$ includes the onsite intra (inter) orbital Coulomb interaction $U$ ($U'$), the Hund's rule coupling $J$ and the pair hopping energy $J'$. We assume spin rotation-invariant interaction parameters $U = 2.4$ eV, $U' = U/2$, and $J = J' = U/4$. Because of the large bandwidth in the collapsed tetragonal phase, these comparatively large values are necessary to bring the system close to the RPA instability. Note however, that the symmetry of the superconducting gap in this system does not change, even if significantly reduced parameter values are considered.

The parameter values used in the RPA method are renormalized with respect to those in the DFT+DMFT method, because the electronic self-energy is neglected in the usual RPA scheme[1,2].

## II. THREE DIMENSIONAL FERMI SURFACE

We extracted the three-dimensional Fermi surface of collapsed tetragonal CaFe$_2$As$_2$ and KFe$_2$As$_2$ from FPLO using $50 \times 50 \times 25$ k-point grids in the two-Fe equivalent Brillouin zone (Fig. 1). For CaFe$_2$As$_2$ only electron pockets in the Brillouin zone corners are observed. The three-dimensional hole pocket that arises from the bands at M $(\pi, \pi, 0)$ is so small that it is not detected here with the given k-resolution. In KFe$_2$As$_2$ one of the hole cylinders is highly dispersive, while the central hole cylinder and the electron pockets located in the corners are nested throughout the entire Brillouin zone.

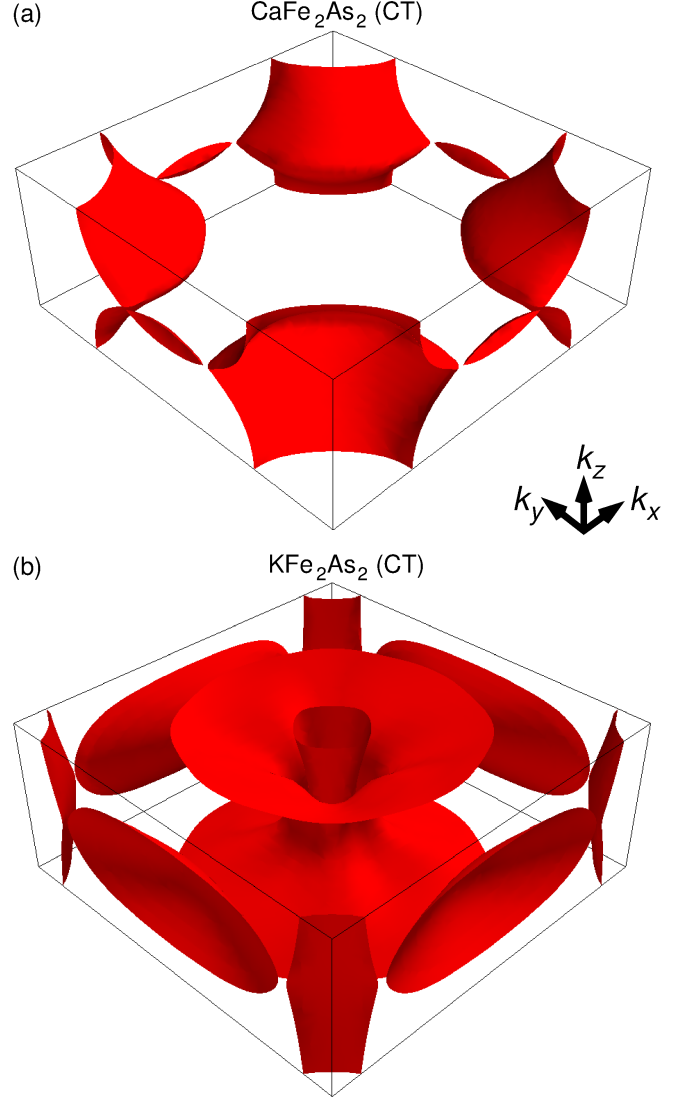

FIG. 1. Three-dimensional Fermi surface of (a) collapsed tetragonal CaFe$_2$As$_2$ and (b) collapsed tetragonal KFe$_2$As$_2$ (both at $P = 21$ GPa) shown in the two-Fe equivalent Brillouin zone. The $\Gamma$-point is located in the center of the displayed volume.

## III. DFT BANDSTRUCTURE AND FERMI SURFACE OF THE NON-COLLAPSED PHASE UNDER PRESSURE

To verify that electron pockets in KFe$_2$As$_2$ do not grow continously with applied pressure, but rather appear as a

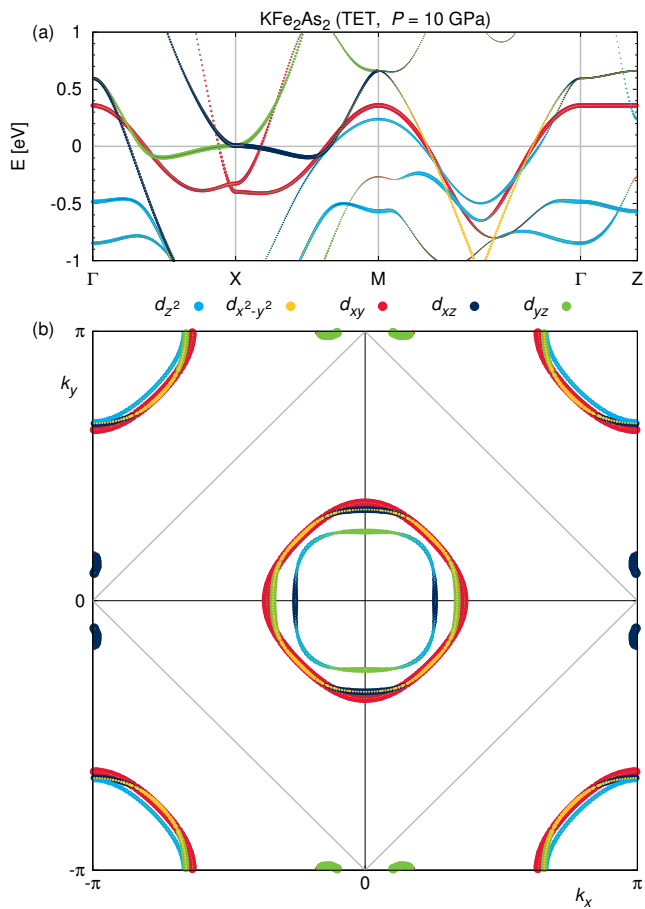

FIG. 2. (a) Electronic bandstructure in the one-Fe equivalent Brillouin zone and (b) Fermi surface of the $P \sim 10$ GPa non-collapsed tetragonal (TET) phase of KFe$_2$As$_2$ at $k_z = 0$. The full plot of the Fermi surface spans the one-Fe equivalent Brillouin zone, while the area enclosed by the grey lines is the two-Fe equivalent Brillouin zone. The colors indicate the weights of Fe $3d$ states.

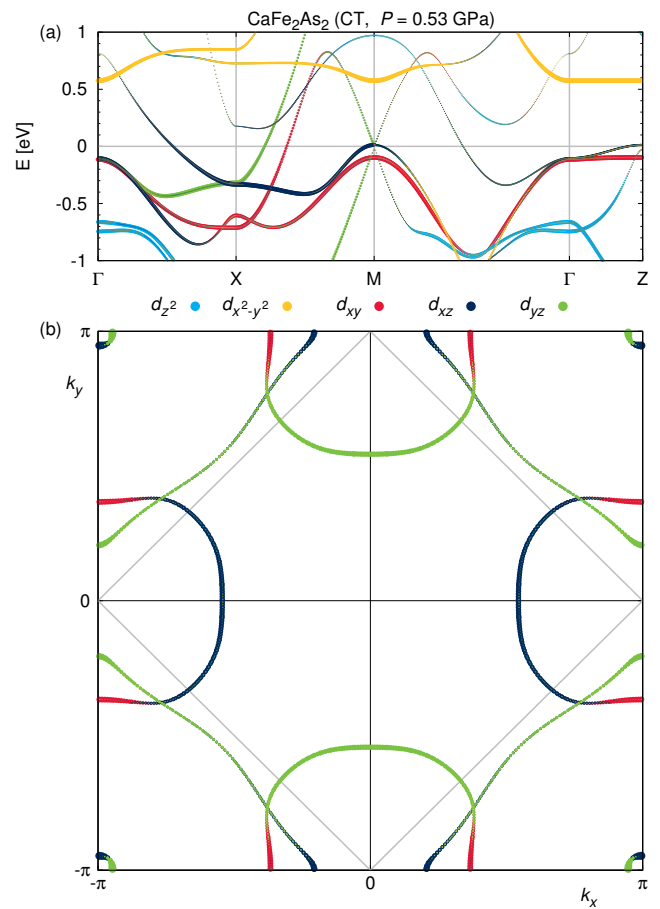

FIG. 3. (a) Electronic bandstructure in the one-Fe equivalent Brillouin zone and (b) Fermi surface of the $P = 0.53$ GPa collapsed tetragonal (CT) phase of CaFe$_2$As$_2$ at $k_z = 0$. The full plot of the Fermi surface spans the one-Fe equivalent Brillouin zone, while the area enclosed by the grey lines is the two-Fe equivalent Brillouin zone. The colors indicate the weights of Fe $3d$ states.

result of the structural collapse, we have also prepared a crystal structure at $P \sim 10$ GPa, using the data from Ref. 3. The lattice parameters were chosen as $a = 3.663$ Å and $c = 13.0$ Å. The fractional arsenic $z$-position ($z_{As} = 0.35750$) was again determined *ab-initio* via structural relaxation using the FPLO code.

No qualitative changes in the bandstructure and Fermi surface (Fig. 2) are observed compared to the ambient pressure structure[4]. Most importantly, electron pockets which are nested with the hole pockets in the Brillouin zone center are not present, even at high pressures slightly below the transition from the tetragonal to the collapsed tetragonal phase. We conclude that the appearance of electron pockets is directly linked to the structural collapse.

## IV. BANDSTRUCTURE, FERMI SURFACE AND SUSCEPTIBILITY AT LOW PRESSURES

In the main paper we used crystal structures at $P = 21$ GPa for both materials of interest to ensure that results are comparable. Here we show the electronic bandstructure, Fermi surface and static susceptibility of CaFe$_2$As$_2$ at low pressure ($P = 0.53$ GPa), right after the structural collapse. We used the experimental structural parameters from Ref. 5, which are the same as in our previous GGA+DMFT study[6]. The differences compared to the high pressure results in the main text are purely quantitative.

The electronic bandstructure and Fermi surface are shown in Fig. 3. The static susceptibility is shown in Fig. 4. The overall enhancement of the susceptibility compared to the figures in the main text is a consequence of the smaller bandwidth at low pressure. The particular enhancement of the peak between X and M in the $d_{xz}$ di-

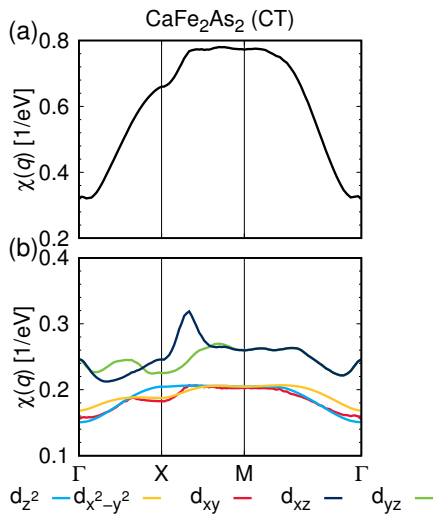

FIG. 4. Summed static susceptibility (a) and its diagonal components $\chi_{aa}^{aa}$ (b) in the eight-band tight-binding model for CaFe$_2$As$_2$ (CT, $P = 0.53$ GPa) in the one-Fe Brillouin zone. The colors identify the Fe 3d states.

agonal element of the susceptibility has its origin in the hole band crossing the Fermi level at the M point, which in turn contributes a small three-dimensional pocket.

## V. DETAILS OF THE LDA+DMFT CALCULATION

For our DMFT calculations we used a temperature of 290 K, i.e. an inverse temperature $\beta = 40$ eV$^{-1}$. The interaction parameters are defined in terms of Slater integrals[7] with an on-site Coulomb interaction $U = 4$ eV and Hund's rule coupling $J = 0.8$ eV. As the double counting correction we used the fully localized limit[8,9] (FLL) scheme. Before the continuation of the imaginary-frequency self-energy to the real axis by stochastic analytic continuation[10], the calculation was converged with $2 \times 10^7$ Monte-Carlo sweeps for the solution of the impurity model.

From the LDA+DMFT calculation we obtain the spectral function on real frequencies from the analytically continued self-energy as

$$A_{vv'}(k, \omega) = -\frac{1}{\pi} \Im \left[ \frac{1}{(\omega + \mu)\delta_{vv'} - \epsilon_{vv'}(k) - \Sigma_{vv'}(k, \omega)} \right], \quad (1)$$

where $\mu$ is the chemical potential, $\epsilon_{vv'}(k) = \epsilon_v(k)\delta_{vv'}$ are the eigenenergies of the LDA Hamiltonian and $\Sigma_{vv'}(k, \omega)$ is the impurity self-energy upfolded to Bloch space by projectors $P_{mv}(k)$ (see Ref. 11 and 12)

$$\Sigma_{vv'}(k, \omega) = \sum_m P_{mv}^\dagger(k) \left( \Sigma_{mm}(\omega) - \Sigma_{DC} \right) P_{mv'}(k), \quad (2)$$

with the double-counting correction term $\Sigma_{DC}$ taken in the fully-localized limit (FLL)[8,9]. The diagonal entries

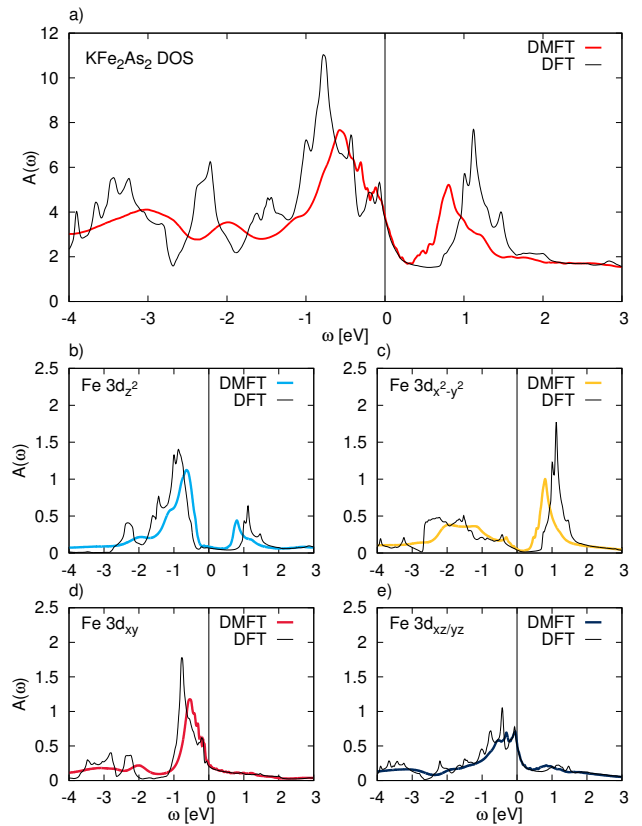

FIG. 5. The k-integrated spectral function (DOS) of KFe$_2$As$_2$ at $P = 21$ GPa as obtained within DFT compared to LDA+DMFT. The upper panel (a) shows the total DOS, with the DFT result (black line) and the renormalized LDA+DMFT result (red line). The lower four panels show the partial DOS of the Fe 3d orbitals: (b) $3d_{z^2}$, (c) $3d_{x^2-y^2}$, (d) $3d_{xy}$ and (e) $3d_{xz/yz}$ orbital.

of this object are then integrated over all bands $v$ and k-points $k$ to obtain the total density of states within LDA+DMFT shown in Fig. 5 (a). Additional projection onto the Fe 3d orbitals (denoted by index $m$) by

$$A_{mm'}(k, \omega) =$$
$$-\frac{1}{\pi} \Im \left[ \sum_{vv'} P_{mv}(k) \frac{1}{(\omega + \mu)\delta_{vv'} - \epsilon_{vv'}(k) - \Sigma_{vv'}(k, \omega)} P_{m'v'}^\dagger(k) \right], \quad (3)$$

is then used in the same fashion to obtain the orbital resolved density of states shown in Fig. 5 (b)-(e).

The k-resolved spectral function shown in Fig. 6 was obtained in the same manner, but only integrating over all bands $v$ in order to keep the k-dependence. The orbital resolved Fermi surfaces shown in Fig. 7 and Fig. 8 were obtained by evaluating $A_{mm}(\omega)$ at the Fermi energy ($\omega = \omega_F$) on a $200 \times 200$ k-grid in the two-Fe Brillouin zone.

We observe moderate effects of renormalization and broadening in the density of states (see Fig. 5) with

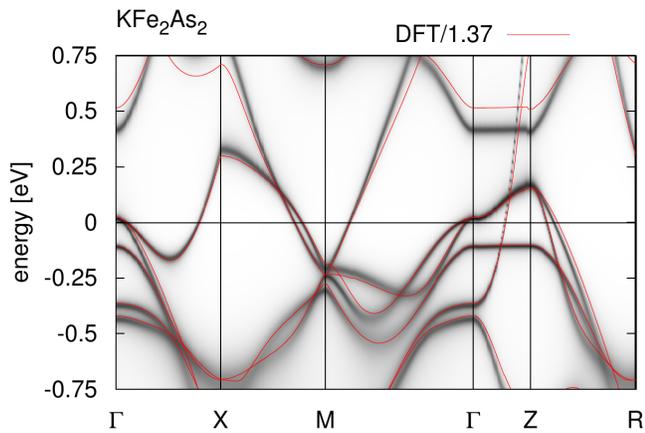

FIG. 6. The k-resolved spectral function of KFe$_2$As$_2$ at $P = 21$ GPa as obtained within LDA+DMFT. The red lines show the DFT bandstructure for comparison, rescaled by the average mass enhancement of 1.37. The labels on the $x$-axis correspond to high-symmetry points in the two-Fe Brillouin zone.

no significant change at the Fermi level. This is confirmed by the k-resolved spectral function (see Fig. 6). Using the FLL double-counting correction, we do not observe any topological changes in the Fermi surface, which closely resembles the DFT result (see Fig. 7 and Fig. 8). However, we observe a closing of the pockets at $\Gamma$ when the nominal double-counting correction[13] is considered. This indicates that the LDA+DMFT results slightly depend on the double-counting. Note that these changes do not affect our conclusions regarding the superconducting pairing symmetry since only a small fraction of the Brillouin zone around $k_z = 0$ is affected. As the relevant electron-hole nesting takes place around $k_z = \pi$, our superconducting pairing calculations based on the DFT bandstructure should be robust with respect to the effects of correlations.

Note that the Fermi surface plots (Fig. 7 and Fig. 8) have to be rotated by 45° in the plane in order to compare with the Fermi surface plots in the original paper. In 122 compounds the Z-point of the two-Fe Brillouin zone corresponds to the M-point of the one-Fe Brillouin zone.

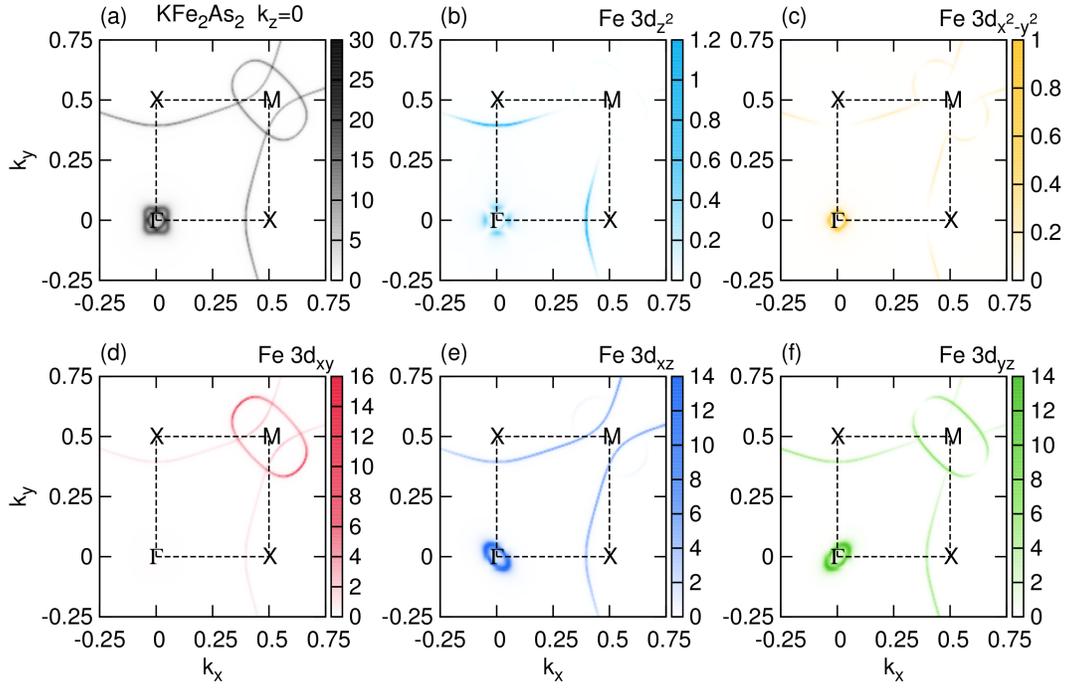

FIG. 7. The $k_z = 0$ Fermi surface of KFe$_2$As$_2$ at $P = 21$ GPa as obtained within LDA+DMFT in the two-Fe Brillouin zone. Panel (a) shows the total spectral function, while panels (b)-(f) show the orbital resolved spectral function for the Fe $3d$ orbitals.

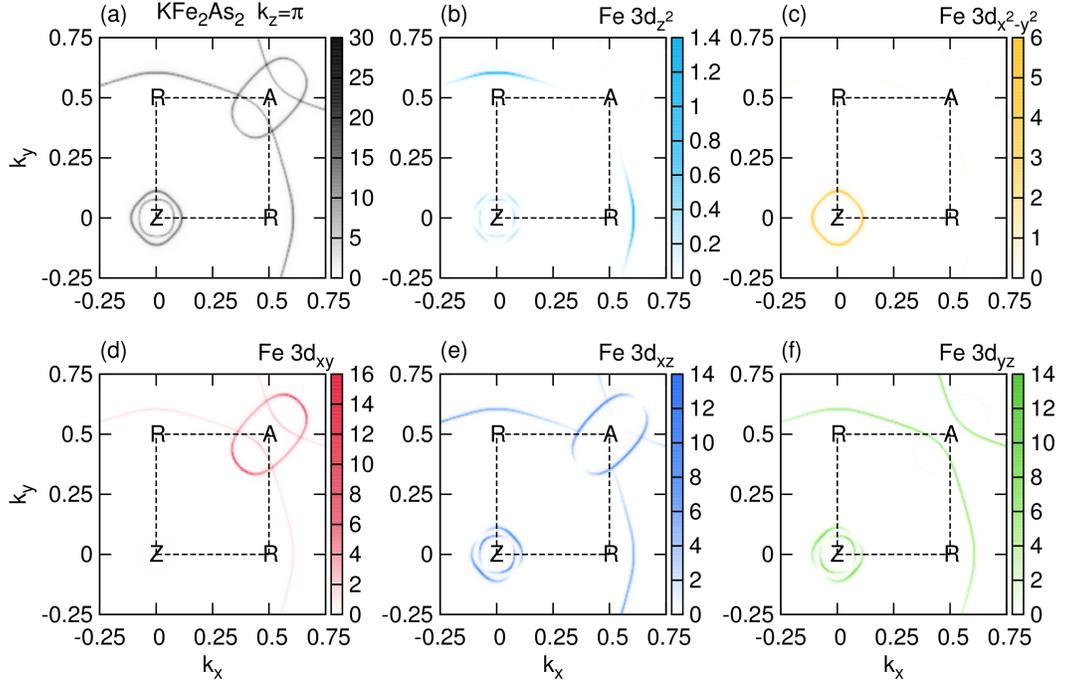

FIG. 8. The $k_z = \pi$ Fermi surface of KFe$_2$As$_2$ at $P = 21$ GPa as obtained within LDA+DMFT in the two-Fe Brillouin zone. Panel (a) shows the total spectral function, while panels (b)-(f) show the orbital resolved spectral function for the Fe $3d$ orbitals.



\end{document}